\documentclass[twocolumn,aps,pre,floatfix,showpacs,amsmath,amssymb]{revtex4}

\usepackage{epsfig}
\usepackage{graphicx}
\usepackage{dcolumn}
\usepackage{bm}
\usepackage[colorlinks=true,dvipdfm]{hyperref}

\usepackage{multirow}
\usepackage{booktabs}

\begin{document}
\title{Generalized dynamic scaling for quantum critical relaxation in imaginary time}
\author{Shuyi Zhang}
\author{Shuai Yin} \email{sysuyinshuai@gmail.com}
\author{Fan Zhong}

\affiliation{State Key Laboratory of Optoelectronic Materials and Technologies, School of Physics and Engineering, Sun Yat-sen University, Guangzhou 510275, People's Republic of China}

\begin{abstract}

We study the imaginary-time relaxation critical dynamics of a quantum system with a vanishing initial correlation length and an arbitrary initial order parameter $M_0$. We find that in quantum critical dynamics, the behavior of $M_0$ under scale transformations deviates from a simple power-law, which was proposed for very small $M_0$ previously. A universal characteristic function is then suggested to describe the rescaled initial magnetization, similar to classical critical dynamics. This characteristic function is shown to be able to describe the quantum critical dynamics in both short- and long-time stages of the evolution. The one-dimensional transverse-field Ising model is employed to numerically determine the specific form of the characteristic function. We demonstrate that it is applicable as long as the system is in the vicinity of the quantum critical point. The universality of the characteristic function is confirmed by numerical simulations of models belonging to the same universality class.

\end{abstract}


\maketitle

\date{\today}

\section{\label{intro}Introduction}
Divergent properties of a system near its critical point are usually characterized by the critical exponents \cite{Wilson}, describing the dependence of macroscopic quantities on the deviation from the critical point. For example, at a distance $g$ from the critical point, the order parameter $M$ behaves as $M\sim g^{\beta}$, the correlation length $\xi$ as $\xi\sim g^{-\nu}$, and the correlation time $\zeta \sim \xi^z$, where $\beta$ and $\nu$ are the static exponents and $z$ is the dynamic exponent. Although the number of microscopic degrees of freedom is huge, the number of independent critical exponents is small, as these critical exponents satisfy several scaling laws. Whether the power laws with the several exponents are enough to describe universal critical properties is an interesting and fundamental question in both classical \cite{Stanley,Ma} and quantum phase transitions \cite{Sachdevbook}.

In this paper, we study the relaxation quantum critical dynamics. Theoretical studies on the quantum critical dynamics have been partly stimulated by developments in experimental technologies \cite{Greiner,ulm,pyka,Meinert}, which provide effective platforms to manipulate quantum many-body systems and detect their nonequilibrium quantum phenomena. Although a lot of effort has been devoted to understanding the universal dynamic properties \cite{Sachdevbook,Dziarmaga,polrmp}, long-range entanglement and nonequilibrium nature make this issue difficult to tackle in both analytical and numerical aspects \cite{sachdevpt}.

In classical critical dynamics \cite{Hohenberg}, the dynamic exponent and the static ones are decoupled. In quantum case, it is well known that there is a mapping between a $d$-dimensional quantum system and a corresponding $(d+1)$-dimensional classical system. The additional dimension comes from the inverse of the temperature, and plays the role of the imaginary time. As the imaginary time has an identical dimension to the real time, parts of the dynamic and static critical properties are intimately intertwined in quantum critical phenomena \cite{Sachdevbook,Sondhi}. For example, hyperscaling scaling laws in quantum criticality include the dynamic exponent $z$ and the static ones together \cite{Sachdevbook,Sondhi}. So, besides the usual static critical exponents and the dynamic one, it is believed that no additional exponent is necessary in quantum critical dynamics, compared to the equilibrium critical phenomena. A typical example is the driving quantum critical dynamics, in which the Kibble-Zurek mechanism \cite{kz} and the finite-time scaling theory \cite{fts1,fts2} show that nonequilibrium critical properties are well described by the exponents in the equilibrium quantum critical phenomena, even at finite temperatures \cite{fts2}.

However, it has been discovered that analogous to the case of classical short-time dynamics \cite{Janssen,Lijpa,Liprl,Zhengrev}, an additional critical exponent $\theta$ is needed to describe the universal critical initial slip in the imaginary-time evolution near a quantum critical point \cite{yinprb}. Starting with a state with vanishing initial correlations and small initial order parameters $M_0$, after a microscopic stage, a system enters the critical initial slip stage, in which $M$ increases with the imaginary time $\tau$ as $M\propto M_0\tau^\theta$. This behavior lasts until a crossover time $\tau_\textrm{cr}$, characterised by $\tau_\textrm{cr}\sim M_0^{-x_0/z}$ with $x_0$ being the scaling dimension of $M_0$ and satisfying $x_0=\theta z+\beta /\nu$. For $\tau> \tau_\textrm{cr}$, $M$ decays according to $M\sim\tau^{-\beta/\nu z}$. Scaling behaviors in both short and long times are shown to be described by unified scaling forms, which can be used to determine the critical point and critical exponents in the short-time stage. The advantage of this method is that it overcomes the critical slowing down and the divergence of the entanglement entropy~\cite{yinprb}.

The initial magnetization $M_0$ plays an essential role in characterizing the initial condition in classical \cite{Janssen,Lijpa,Liprl,Zhengrev} and quantum short-time critical dynamics \cite{yinprb}. Both $M_0=0$ and the saturated $M_0$ are the fixed points of the initial order parameter under scale transformations \cite{Janssen,Lijpa,Liprl,Zhengrev,yinprb}. When evolution begins with $M_0=0$, all the evolution is the initial slip. When evolution begins with the saturated $M_0$, there is no critical initial slip. When $M_0$ is very small and thus close to the fixed point $M_0=0$, the rescaled order parameter, $M_0^{'}$, under a scale transformation with a rescaling factor $b$, is $M_0^{'}=b^{x_0}M_0$. In between, for an arbitrary $M_0$, in classical critical dynamics, Zheng \cite{Zhengprl} showed that universal behavior also exists, however, $b^{x_0}M_0$ cannot characterize the rescaled order parameter $M_0^{'}$ for large $M_0$. He proposed that $M_0^{'}$ can be represented by a {\it universal characteristic function}, with $b$ and $M_0$ being its arguments \cite{Zhengprl}. Numerical results have confirmed this proposal \cite{Zhengprl}. Moreover, the universality of the characteristic function has been verified in other classical systems \cite{Cheneur,Lipre,fangijmp,heijmp,chenpa}. A natural question is then does quantum imaginary-time evolution also need such a universal characteristic function?

In this paper, we show that the power-law scale transformation of $b^{x_0}M_0$ is not enough to characterize the rescaled initial magnetization in the imaginary-time relaxation near the quantum critical point with an arbitrary $M_0$. We suggest that a universal characteristic function $U(b,M_0)$ can be introduced, as in the classical case \cite{Zhengprl,Cheneur,Lipre,fangijmp,heijmp,chenpa}, to characterize the behavior of an arbitrary initial parameter $M_0$ under the scale transformation. Both the short- and the long-time critical dynamics are well described by the generalized scale transformation containing this function as the rescaled initial order parameter. The characteristic function is shown to be determined by the scaling factor $b$ and the original $M_0$, and independent of $g$ and the symmetry-breaking field $h$. When $M_0$ is small, $U(b,M_0)$ recovers the power-law form $b^{x_0}M_0$, while for the saturated $M_0$ we have $U(b,M_0)=M_0$. This function is universal, since it is identical for different models belonging to one universality class. These conclusions are verified by numerical results of the order parameter and the entanglement entropy in the one-dimensional ($1$D) transverse-field Ising model and quantum XXZ model and the transverse-field Ising-ladder model.

The rest of the paper is organized as follows. In Sec.~\ref{imevo}, we illustrate the imaginary-time Schr\"{o}dinger equation and compare it with the classical master equation. Then, in Sec.~\ref{scalingt}, we discuss the scale transformations containing the universal characteristic function $U(b,M_0)$ and its properties. The scaling theory is verified in Sec.~\ref{ver}. Firstly, a mean-field theory is developed to study the scaling properties of the characteristic function in Sec.~\ref{MF}. Then, in Sec.~\ref{scaling}, we determine $U(b,M_0)$ of the quantum Ising chain and confirm its scaling properties numerically. Subsequently, the universality of $U(b,M_0)$ is verified for various realizations of the initial magnetization and for other models in the same universality class. Finally, a summary is given in Sec.~\ref{summary}.

\section{\label{imevo}Imaginary-time quantum evolution}
Studies on the imaginary-time quantum critical dynamics have attracted a lot of attention recently \cite{polprb1,polprb2,poljpcm,Sachdevnp}. Some scaling properties of the imaginary-time quantum critical dynamics can be used to predict the behavior of the real-time critical dynamics \cite{polprb1,polprb2,poljpcm,Sachdevnp,Karrasch}. Moreover, simulations of the imaginary-time dynamics can be readily realized in both quantum Monte Carlo \cite{polprb1,polprb2,poljpcm,binder} and density-matrix renormalization group methods \cite{schollrmp,dmrgim}. More importantly, the imaginary-time quantum dynamics has its own physical realization. For example, it can be regarded as the evolution controlled by a non-Hermitian Hamiltonian with strong dissipation \cite{Moiseyevbook,Nesterov}.

The imaginary-time evolution of a quantum state $|\psi(\tau)\rangle$ is governed by the Schr\"{o}dinger equation with a Hamiltonian $H$ and the time $t$ replaced by $-i\tau$ \cite{Justinbook,Altlandbook},
\begin{equation}
{\frac{\partial}{\partial\tau} |\psi(\tau) \rangle}=-H|\psi(\tau)\rangle, \label{scheq}
\end{equation}
in which $|\psi(\tau)\rangle$ satisfies the normalization condition $\langle \psi(\tau)|\psi(\tau) \rangle=1$. In the real-time evolution, the normalization condition is naturally satisfied because of the unitarity of the evolution operator. For the imaginary-time evolution, the normalization condition has to be imposed on \cite{Justinbook}. The formal solution to Eq. (\ref{scheq}) is
\begin{equation}
|\psi(\tau) \rangle=Z^{-1}\textrm{exp}(-H\tau)|\psi_0 \rangle, \label{sscheq}
\end{equation}
where $|\psi_0 \rangle\equiv |\psi(0) \rangle$ is the initial state, and
\begin{equation}
Z=\|\textrm{exp}(-H\tau)|\psi_0\rangle\|, \label{normcon}
\end{equation}
is the normalization factor with $\|\centerdot\|$ denoting a modulo operation.

In order to compare the imaginary-time quantum dynamics with the classical thermal dynamics, we now derive an equation describing the evolution of the normalized wave function, $|\phi(\tau)\rangle\equiv Z^{-1}|\psi(\tau) \rangle$. Substituting Eqs.~({\ref{sscheq}}) and (\ref{normcon}) into Eq.~(\ref{scheq}), we obtain
\begin{equation}
{\frac{\partial}{\partial\tau} |\phi(\tau) \rangle}=-H|\phi(\tau)\rangle+\bar{E}(\tau)|\phi(\tau)\rangle, \label{scheqn}
\end{equation}
where $\bar{E}(\tau)=\langle \phi(\tau)|H|\phi(\tau)\rangle$ is the averaged energy. Expanding $|\phi(\tau)\rangle$ in the eigenstates of the Hamiltonian,
\begin{equation}
|\phi(\tau) \rangle=\sum_i C_i(\tau)|E_i\rangle, \label{scheqn1}
\end{equation}
we arrive at the evolution of the probability $P_i(\tau)\equiv |C_i(\tau)|^2$ of finding the $i$th eigenstate $|E_i\rangle$ with the eigenvalue $E_i$,
\begin{equation}
{\frac{\partial P_i(\tau)}{\partial\tau}}=-\left[E_i-\bar{E}(\tau)\right]P_i(\tau). \label{scheqn2}
\end{equation}
Two remarks are in order here. (a) Equation (\ref{scheqn2}) is a typical dissipation equation. If $E_i>\bar{E}(\tau)$, $P_i$ will decay; if $E_i<\bar{E}(\tau)$, on the other hand, $P_i$ will increase. The steady solution of this equation corresponds to $\bar{E}=E_i$. (b) For a system with a first exciting gap $\Delta=E_1-E_0$, the ground-state energy $E_0$ is always smaller than $\bar{E}(\tau)$. Thus the system will tend to its ground state after a typical time scale $\zeta_\tau\sim \Delta^{-1}$. Hence the imaginary-time evolution is a commonly used method to find the ground state \cite{dmrgim}.

The reason for the similarity between the imaginary-time quantum critical dynamics and the classical critical dynamics can be inspected by comparing Eq.~(\ref{scheqn2}) with the classical master equation \cite{binder},
\begin{equation}
{\frac{\partial P_i(t)}{\partial t}}=\sum_j\left[W_{j\rightarrow i}P_j(t)-W_{i\rightarrow j}P_i(t)\right], \label{ccheqn}
\end{equation}
in which $W_{j\rightarrow i}$ is the transition probability from the $j$th to the $i$th state. $W_{j\rightarrow i}$ must fulfill the detailed balance condition, which is $W_{j\rightarrow i}/W_{i\rightarrow j}=\textrm{exp}\left[-(E_i-E_j)/T\right]$ with $T$ being the temperature. Both equations describe a dissipation process. The probability of the high-energy excitations decays fast with the time evolution, whereas the low-energy modes, controlling the critical phenomena, are left over. As a result, Eqs.~(\ref{ccheqn}) and (\ref{scheqn2}) exhibit similar evolution properties, especially the critical initial slip behavior in the short-time stage. However, the critical dynamics described by Eq.~(\ref{scheqn2}) is essentially different from that described by the classical master equation (\ref{ccheqn}), even for two models in the same static universality class. For example, the dynamic exponent corresponding to Eq.~(\ref{scheqn2}) is $z=1$ for the $1$D quantum Ising model \cite{Sachdevbook}, while that to Eq. (\ref{ccheqn}) is $z\simeq2.1667(5)$ for the $2$D classical Ising model \cite{Night,Albano}. The difference between them also explains the fact that the initial-slip exponent $\theta$ is different from the classical one \cite{yinprb}.

\section{\label{scalingt}Generalized short-time critical dynamics}
In classical short-time critical dynamics, a universal characteristic function has been proposed to describe the rescaled initial order parameter in the scale transformation characterizing evolution with an arbitrary initial magnetization \cite{Zhengprl}. In this section, we suggest a similar scale transformation which contains a universal characteristic function to describe the imaginary-time evolution starting with an arbitrary initial magnetization in quantum critical dynamics.

Universal behavior near the quantum critical point is controlled by the low-lying energy levels. Because of the dissipative nature of Eq.~(\ref{scheq}), when a quantum system is quenched to the vicinity of its critical point, contributions from the high-energy levels decay fast in the microscopic initial stage. After this stage, the critical system enters the universal short-time stage. For both very small $M_0$ and the saturated $M_0$, previous studies \cite{yinprb} have shown that a scaling form connects the scaling behavior in this stage to that in the long-time stage. Thus, it is justified to expect that criticality should also exist for an arbitrary $M_0$. However, as $M_0$ is not in the vicinity of its fixed point, a simple power-law relation is not enough to describe the universal behavior. For example, if the relation $M_0^{'}=b^{x_0}M_0$ were right for all $M_0$, $M_0^{'}$ would be larger than the saturated value, which is clearly not physical. Inspired by the idea of the universal characteristic function applied to the classical case \cite{Zhengprl}, we suggest that the universal characteristic function $U(b,M_0)$ is the rescaled initial order parameter $M_0^{'}$ for the rescaling factor $b$.

Accordingly, in case of the order parameter, we have
\begin{equation}
M(\tau,g,h,M_0)=b^{-\beta/\nu}M(b^{-z}\tau,b^{1/\nu}g,b^{\beta\delta/\nu}h,U(b,M_0)), \label{gsc}
\end{equation}
where $U(b,M_0)$ is the universal characteristic function \cite{Zhengprl}, and $\delta$ is another static critical exponent defined as $M\propto h^{1/\delta}$ at $g=0$.

It should be noted that $U$ is to be included in the scale transformations for all macroscopic physical quantities. Besides the order parameter, we can also measure the evolution of the entanglement. As a unique physical quantity of a quantum system, the entanglement is usually measured as $S=-\textrm{Tr}(\rho\textrm{log}\rho)$ if we apply the definition the von-Neumann entropy, where $\rho$ is the reduced density matrix of half of the system \cite{Eisert,Amico,osterloh} and the base of logarithm is $2$ throughout. For a $1$D system near its critical point $S=(c/6)\textrm{log}\xi$ \cite{Eisert,Amico,osterloh}, with $c$ being the central charge. Using
the characteristic function, we can write down the generalized scale transformation of the correlation length,
\begin{equation}
\xi(\tau,g,h,M_0)=b\xi(b^{-z}\tau,b^{1/\nu}g,b^{\beta\delta/\nu}h,U(b,M_0)). \label{xisc}
\end{equation}
Therefore, the entanglement entropy $S$ satisfies
\begin{equation}
S(\tau,g,h,M_0)=\frac{c}{6}\textrm{log}b+S(b^{-z}\tau,b^{1/\nu}g,b^{\beta\delta/\nu}h,U(b,M_0)). \label{eesc}
\end{equation}

Here are some properties of the function $U(b,M_0)$. (a) At both fixed points of $M_0$, i.e., $M_0=0$ or the saturated one, we have $U(b,M_0)=M_0$ for any given $b$. (b) The value $U(b,M_0)$ depends only on $M_0$ and the rescaling factor $b$, but not on any other parameters, $g$ and $h$, for example.
In other words, Eq.~(\ref{gsc}) always holds as long as the system is near its critical point. (c) When the initial order parameter is small, $M_0\rightarrow0$, the characteristic function returns to a simple power-law relation $U(b,M_0)\rightarrow b^{x_0}M_0$. (d) Exactly at the critical point $g=0$ and $h=0$, by setting $b=\tau^{1/z}$, Eq.~(\ref{gsc}) becomes
\begin{equation}
M(\tau,M_0)=\tau^{-\beta/\nu z}f_M(U(\tau^{1/z},M_0)), \label{gscf}
\end{equation}
where $f_M$ is the scaling function related to $M$. In the long-time stage, any information of the starting $M_0$ should be ``forgotten", except for the sign of $M_0$. Hence we have $f_M(U(\tau^{1/z},M_0))\sim \textrm{sgn}(M_0)$ for any $M_0$ when $\tau\rightarrow \infty$, where $\textrm{sgn}$ is the sign function. In the short-time stage, $\tau\ll M_0^{-z/x_0}$, the scaling behavior for very small $M_0$ can be easily restored \cite{yinprb}. As $U(b,M_0)\rightarrow b^{x_0}M_0$, we have $f_M(U(\tau^{1/z},M_0))\sim M_0\tau^{x_0/z}$ and $M$ increases with $\tau$ as $M\sim M_0\tau^\theta$ \cite{yinprb}, as in the classical situation \cite{Janssen,Lijpa,Zhengprl}. Similar to the case of $M$, let $b=\tau^{1/z}$, and the generalized scaling form for $S$ reads
\begin{equation}
S(\tau,M_0)=\frac{c}{6z}\textrm{log}\tau+f_S(U(\tau^{1/z},M_0)). \label{xiscf}
\end{equation}
We define an entanglement entropy-difference, $\Delta S$, by
\begin{equation}
\Delta S(\tau, M_0) \equiv f_S(U(\tau^{1/z},M_0))-f_S(U(\tau^{1/z},0)). \label{eescf}
\end{equation}
For long times, if $M_0\neq 0$, $f_S(U(\tau^{1/z},M_0))$ tends to a constant, $\Delta S(\infty, 1)$, independent of $M_0$ \cite{yinprb}, whereas in the short times and for a small $M_0$, we restore \cite{yinprb},
\begin{equation}
\Delta S(\tau, M_0)\propto M_0^2\tau^{2x_0/z}. \label{eescf1}
\end{equation}

\section{\label{ver}Verification of the generalized scale transformations}
In this section, we verify the scale transformation proposed for the universal imaginary-time quantum critical dynamics with an arbitrary initial magnetization $M_0$ and determine the universal characteristic function $U(b,M_0)$. The $1$D transverse-field Ising model is taken as an example. First, a mean-field theory is developed to show the universal behavior for an arbitrary initial order parameter. An analytic result of the characteristic function is obtained in this mean-field approximation. Second, we shall confirm Eq.~(\ref{gsc}) at $g=0$ and $h=0$ and determine $U(b,M_0)$. Third, we show that $U(b,M_0)$ is independent of $g$ and $h$. Finally, the universal properties are confirmed by examining $U$ for various initial states and for different models in the universality class of the quantum Ising chain and comparing it with the quantum Ising chain.

\subsection{\label{monum}Model, numerical method, and initial state}
In the following, we mainly use the $1$D transverse field Ising model. The Hamiltonian reads
\begin{equation}
H_I=-\sum\limits_{n}\sigma_n^z\sigma_{n+1}^z-h_x\sum\limits_{n}\sigma_n ^x-h\sum\limits_{n}\sigma_n ^z,
\label{HIsing}
\end{equation}
where $\sigma_n^x$ and $\sigma_n^z$ are the Pauli matrices in $x$ and $z$ direction, respectively,
at site $n$, $h_x$ is the transverse field and $h$ is the symmetry-breaking field. We have set the
Ising coupling to unity as our energy unit. The order parameter is defined
as $M=\langle\sigma^z_n\rangle$, where the angle brackets denote the quantum average of the operator over
each site. The critical point of model~(\ref{HIsing}) is
$h_{xc}=1$ and $h=0$. The exact critical exponents are $\beta=1/8$, $\nu=1$, $\delta=15$ and $z=1$
\cite{Sachdevbook}, and the central charge $c=1/2$ \cite{Eisert,Amico}. The critical initial-slip exponent $\theta$ is estimated to be $\theta=0.373$ \cite{yinprb}. This model is realized in CoNb$_2$O$_6$ experimentally \cite{Coldea}.

In order to show the universality of the characteristic function, we also employ the quantum XXZ model in a transverse field with the Hamiltonian \cite{scholl}
\begin{equation}
\begin{split}
H_{XXZ}=&-\sum\limits_{n}\sigma_n^z\sigma_{n+1}^z-J_{xy}\sum\limits_{n}\sigma_n^x\sigma_{n+1}^x
\\&-J_{xy}\sum\limits_{n}\sigma_n^y\sigma_{n+1}^y-h_x\sum\limits_{n}\sigma_n ^x,
\label{HXXZ}
\end{split}
\end{equation}
where $\sigma^y$ is the pauli matrix in $y$ direction and $J_{xy}$ is the coupling constant in the spin $x$ and $y$ directions. For $J_{xy}=0.2$ and $0.3$, the critical points from the ferromagnetic phase to the paramagnetic phase are $h_{xc}=0.8504$ and $0.7682$, respectively. The critical point is determined by the method of the short-time critical dynamics \cite{yinprb}. This method is also used to estimate the critical exponents, including the initial-slip exponent $\theta$. All of the exponents are identical with those of model (\ref{HIsing}), confirming that the phase transition belongs to the same universality class to model (\ref{HIsing}). This model has been realized in Cs$_2$CoCl$_4$ experimentally \cite{Breunig}.

Another model utilized to verify the universality of the characteristic function is the quantum Ising-ladder model with the Hamiltonian \cite{scholl}
\begin{equation}
\begin{split}
H_L=&-\sum\limits_{n}\sum\limits_{\alpha=1,2}\sigma_{\alpha,n}^z\sigma_{\alpha,n+1}^z-\sum\limits_{n}\sigma_{1,n}^z\sigma_{2,n}^z
\\&-h_x\sum\limits_{n}\sum\limits_{\alpha=1,2}\sigma_{\alpha,n}^x, \label{HIsingl}
\end{split}
\end{equation}
where the first and the second terms are the interactions along the ladder and on the rung, respectively, and $\alpha$ denotes the legs of the ladder. The critical point of this model was determined by finite-time scaling method~\cite{fts1} to be $h_x=1.8323$ \cite{chenzy} and the static critical exponents and the dynamic exponent $z$ determined by the same method show that it belongs to the same universality class to model (\ref{HIsing}) \cite{chenzy}. It has also been shown that the initial-slip exponent, $\theta$, is very close for the two models~\cite{yinprb}.

The infinite time-evolving block decimation (ITEBD) algorithm~\cite{vidali} is used to calculate the imaginary-time evolution in Sec.~\ref{scaling}. A quantum state in $1$D can be represented in a matrix product form via Vidal's decomposition and each site is attached with such a matrix. By taking the translational invariance of an infinite homogeneous condition into account, the ITEBD algorithm represents the matrix product form with repeated matrices in one primitive cell. The evolution of a state then is represented by the updating of these matrices according to the local evolution operators, which are obtained by the Suzuki-Trotter decomposition of $\exp(-H \tau)$. The time interval is chosen as $0.01$ and $100$ states are kept. These values are identical with the previous study \cite{yinprb}. Three decimal places are kept in our results, as the increment of $M_0$, which we choose to determine $U$, is $0.002$.

The initial state with an order parameter $M_0$ is prepared in a direct product state. Both homogeneous and staggered initial states have been used to determine the universal short-time properties in the previous study \cite{yinprb}. For models (\ref{HIsing}) and (\ref{HXXZ}), the initial wave-function is chosen as
\begin{equation}
|\psi_0\rangle_I=\bigotimes\limits_{n}\left[(a_{2n}|\uparrow\rangle+b_{2n}|\downarrow\rangle)(a_{2n+1}|\uparrow\rangle+b_{2n+1}|\downarrow\rangle)\right], \label{istate}
\end{equation}
where $a_n$ and $b_n$ are coefficients of the local state at site $n$, $|\uparrow\rangle$ and $|\downarrow\rangle$ are eigenvectors of $\sigma^z$. For a given initial magnetization $M_0$, $a_{2n}=\sqrt{(1+M_{0A})/2}$, $b_{2n}=\sqrt{(1-M_{0A})/2}$, $a_{2n+1}=\sqrt{(1+M_{0B})/2}$ and $b_{2n+1}=\sqrt{(1-M_{0B})/2}$, and $M_{0A}$ and $M_{0B}$ satisfy $(M_{0A}+M_{0B})/2=M_0$. For the homogeneous initial state, $M_{0A}=M_{0B}$, thus $a_{2n}=a_{2n+1}$ and $b_{2n}=b_{2n+1}$. For the quantum Ising-ladder model (\ref{HIsingl}), the local basis vectors are $|\uparrow^l\uparrow^r\rangle$, $|\uparrow^l\downarrow^r\rangle$, $|\downarrow^l\uparrow^r\rangle$ and $|\downarrow^l\downarrow^r\rangle$, where $l$ and $r$ label the two spins on the same rung. We shall choose the homogenerous initial state for the quantum Ising-ladder and the initial wave-function is
\begin{equation}
|\psi_0\rangle_L=\bigotimes\limits_{n} \left[(a_{n}^l |\uparrow^l\rangle+b_{n}^l |\downarrow^l\rangle)(a_{n}^r |\uparrow^r\rangle+b_{n}^r|\downarrow^r\rangle)\right],
\label{listate}
\end{equation}
where $a_{n}^l=a_{n}^r=\sqrt{(1+M_0)/2}$ and $b_{n}^l=b_{n}^r=\sqrt{(1-M_0)/2}$ for a given $M_0$.

\subsection{\label{MF}Mean-field theory}
In this section, we shall study the mean-field theory of the relaxation dynamics from an arbitrary $M_0$. In this mean-field approximation, the universal behavior is confirmed and an analytic expression for the universal characteristic function is obtained for the universality class of model~(\ref{HIsing}).

\subsubsection{Analytic results of $U$ in the mean-field theory}
The mean-field Hamiltonian of the quantum Ising model~(\ref{HIsing}) is \cite{Chakrabarti,yinprb}
\begin{equation}
\tilde{H}_{\rm MF}=-2M\sigma_z-h_x\sigma_x. \label{mfhamil}
\end{equation}
Its critical point is $h_{xc}^{\rm MF}=2$ and the static critical exponents, $\beta^{\rm MF}=1/2$, $\delta^{\rm MF}=3$ and $\nu^{\rm MF}=1/2$, while the dynamic exponent, $z^{\rm MF}=2$ \cite{Chakrabarti,yinprb} and the critical initial-slip exponents $\theta^{\rm MF}=0$ and $x_0^{\rm MF}=1$ \cite{yinprb}. These exponents satisfy $x_0^{\rm MF}/z^{\rm MF}=1/2$ and $\nu^{\rm MF}z^{\rm MF}=1$ \cite{yinprb}.

According to Eq.~(\ref{mfhamil}), the evolution equation for the order parameter, $M$, in the mean-field approximation, is~\cite{yinprb}
\begin{equation}
\frac{dM}{d\tau}=4M-4M^3-2h_xM\sqrt{1-M^2}. \label{dmfhamil}
\end{equation}
Equation~(\ref{dmfhamil}) can be solved analytically. At the critical point, $h_x=h_{xc}^{\rm MF}=2$, the solution for an arbitrary initial magnetization, $M_0$, fulfills an implicit function,
\begin{equation}
Z(M)=8\tau+Z(M_0), \label{fullsol}
\end{equation}
where $Z(M)$ is
\begin{equation}
Z(M)=Y(M)+\textrm{ln}\left[\frac{1+\sqrt{1-M^2}}{|M|}\right], \label{Zfun}
\end{equation}
with
\begin{equation}
Y(M)=\frac{1+\sqrt{1-M^2}}{M^2}. \label{Yfun}
\end{equation}
\begin{figure}
  \centerline{\epsfig{file=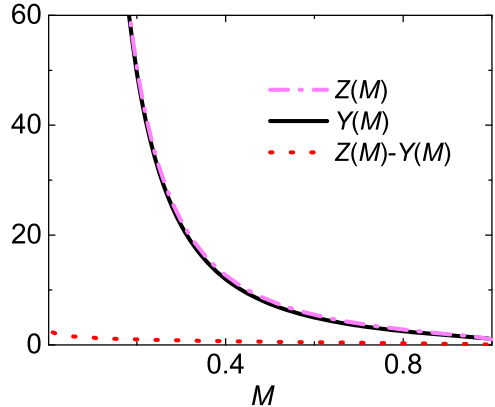,width=1.0\columnwidth}}
  \caption{\label{ZYM} (Color online) Comparison of $Z(M)$, $Y(M)$ and $Z(M)-Y(M)$.}
\end{figure}

Comparing the two parts in the right-hand side of Eq.~(\ref{Zfun}), as displayed in Fig.~\ref{ZYM}, one finds that the first part, $Y(M)$, dominates for $M\leqslant 1$. Therefore, the universal scaling behavior can be well approximated by solving
\begin{equation}
Y(M)=8\tau+Y(M_0). \label{YYfun}
\end{equation}

The analytic explicit solution of Eq.~(\ref{YYfun}) is
\begin{equation}
M(\tau,M_0)={\rm sgn}(M_0)\sqrt{\frac{16\tau+2Y(M_0)-1}{\left[8\tau+Y(M_0)\right]^2}}. \label{SYYfun1}
\end{equation}
In the universal stage, $\tau>1/16$, Eq.~(\ref{SYYfun1}) becomes
\begin{equation}
M(\tau,M_0)\simeq{\rm sgn}(M_0)\sqrt{\frac{2}{Y(M_0)+8\tau}}. \label{SYYfun}
\end{equation}
In the long-time stage, $\tau\gg Y(M_0)$, Eq.~(\ref{SYYfun}) indicates that $M(\tau,M_0)\sim \tau^{-1/2}$. This scaling relation coincides with $M\sim \tau^{-\beta^{\rm MF}/\nu^{\rm MF}z^{\rm MF}}$ by substituting the mean-field critical exponents.

To explore the universal behavior in the short-time stage, we carry out a scale transformation according to Eq.~(\ref{gsc}) with a rescaling factor $b$. Under this scale transformation, Eq.~(\ref{SYYfun}) becomes
\begin{equation}
Mb^{\beta^{\rm MF}/\nu^{\rm MF}}={\rm sgn}(M_0)\sqrt{\frac{2}{Y[U^{\rm MF}(b,M_0)]+8\tau b^{-z^{\rm MF}}}}, \label{SCYYf}
\end{equation}
where $U^{\rm MF}(b,M_0)$ is the universal characteristic function in the mean-field approximation. According to Eq.~(\ref{SCYYf}), $Y[U^{\rm MF}(b,M_0)]$ must satisfy $Y[U^{\rm MF}(b,M_0)]=Y(M_0)b^{-z^{\rm MF}}$. As a result, the universal characteristic function, $U^{\rm MF}$, is solved analytically by
\begin{equation}
\begin{split}
&U^{\rm MF}(b,M_0)=
\\&\sqrt{b^{z^{\rm MF}}\left[2-2\sqrt{1-M_0^2}+b^{z^{\rm MF}}(M_0^2-2+2\sqrt{1-M_0^2})\right]}. \label{SCYYf1}
\end{split}
\end{equation}

\begin{figure}
  \centerline{\epsfig{file=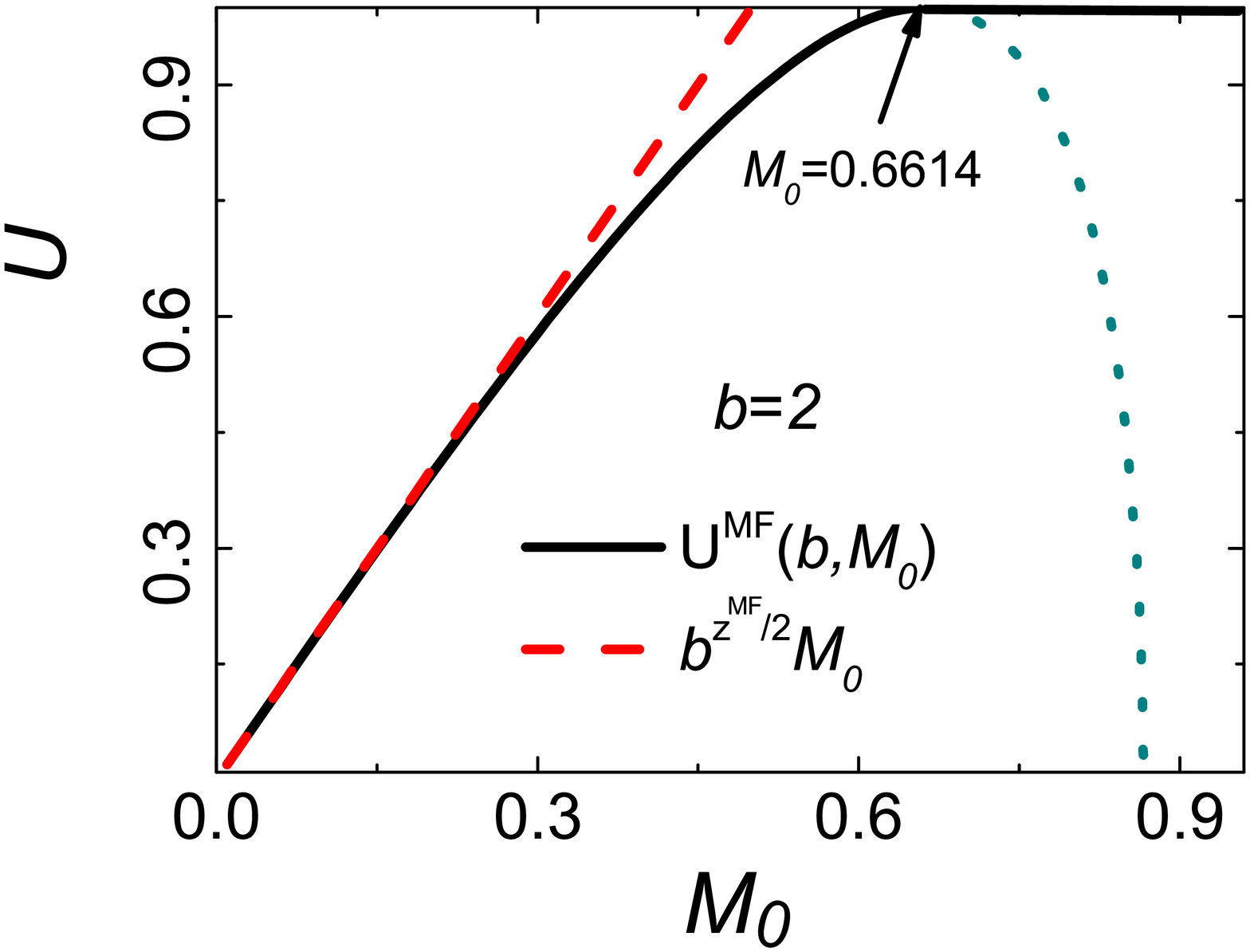,width=0.8\columnwidth}}
  \caption{\label{UMF} (Color online) The universal characteristic function $U^{\rm MF}(b,M_0)$ for $b=2$. At $M_0\simeq 0.6614$, $U^{\rm MF}=1$. The line of $b^{z^{\rm MF}/2}M_0$ is also shown for comparison. The dotted curve shows the non-physical part of Eq.~(\ref{SCYYf1}).}
\end{figure}

However, Eq.~(\ref{SCYYf1}) only describe the scale transformation for $M_0<b^{-z^{\rm MF}}\sqrt{2b^{z^{\rm MF}}-1}$, at which $U^{\rm MF}(b,M_0)$ saturates to $1$. (Fig.~\ref{UMF}). Beyond that, Eq.~(\ref{SCYYf1}) decreases monotonically to $0$ at some $M_0<1$ depending on $b$, unless $b=1$, at which $U^{\rm MF}(1,M_0)=M_0$ as expected. However, this is not a physical result, as $U^{\rm MF}$ should be a monotonically increasing function of $M_0$. Since $U^{\rm MF}(b,1)=1$, we suggest that $U^{\rm MF}(b,M_0)$ is a piecewise function, which is Eq.~(\ref{SCYYf1}) for $M_0<b^{-z^{\rm MF}}\sqrt{2b^{z^{\rm MF}}-1}$, and equals one for $b^{-z^{\rm MF}}\sqrt{2b^{z^{\rm MF}}-1}<M_0<1$.

Some remarks are in order here. (a) The existence of the solution~(\ref{SCYYf1}) to the scale transformation~(\ref{SYYfun}) confirms that the universal scaling behavior exists for an arbitrary initial magnetization, $M_0$. (b) For $M_0\ll 1$, $U^{\rm MF}(b,M_0)=b^{x_0^{\rm MF}}M_0$ with $x_0^{\rm MF}=z^{\rm MF}/2$, restoring the previous result \cite{yinprb}. (c) Equation~(\ref{SCYYf1}) shows that a simple power-law is not enough to describe the scale transformation of the initial order parameter for larger $M_0$. It demonstrates the necessity of introducing the universal characteristic function even in the mean-field theory.

\subsubsection{Verification the scaling properties of $U^{\rm MF}$}
\begin{figure}
  \centerline{\epsfig{file=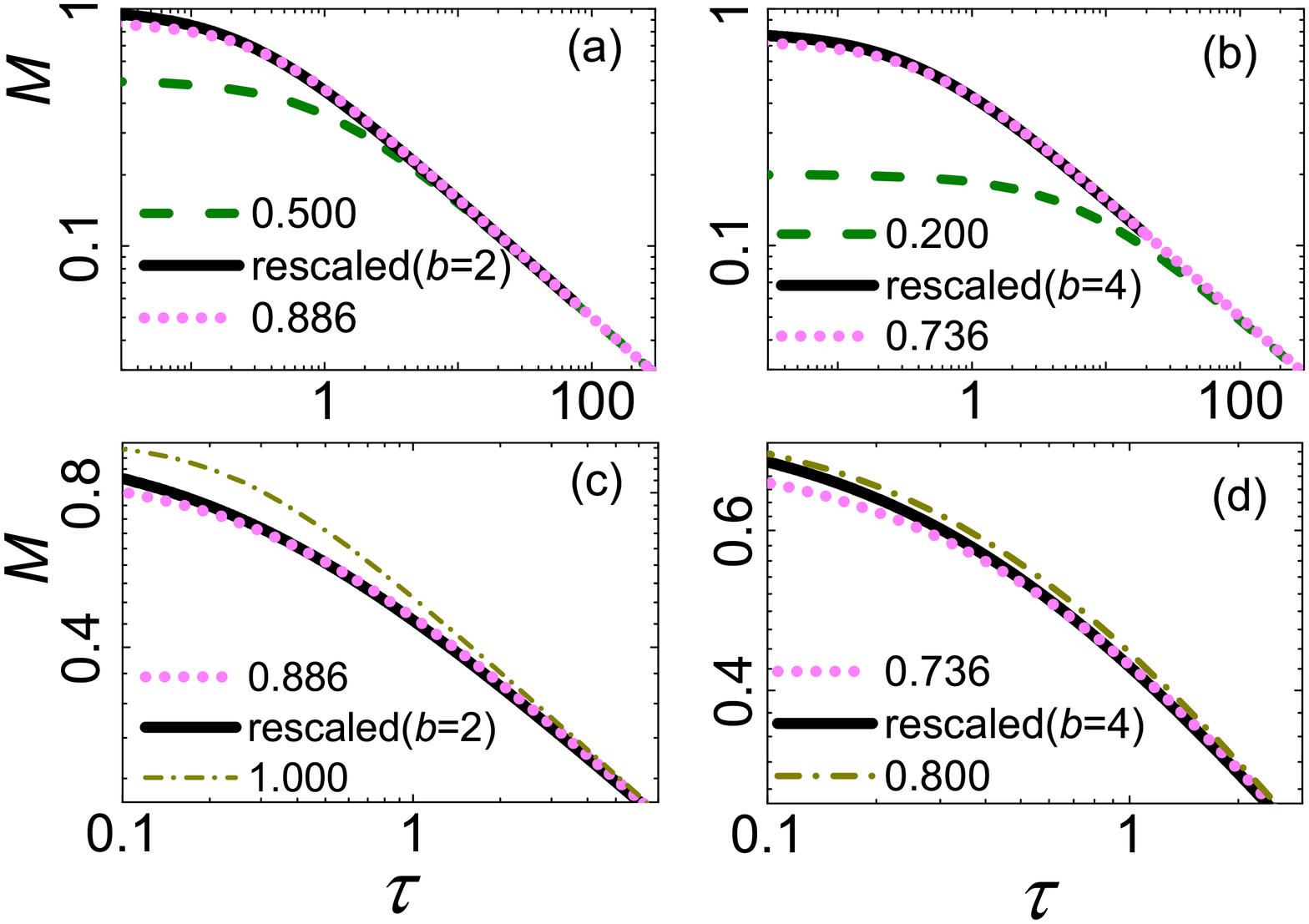,width=1.0\columnwidth}}
  \caption{\label{MFNR} (Color online) Verification of the mean-field solution of the universal characteristic function for (a) $U^{\rm MF}(2,0.5)$ and (b) $U^{\rm MF}(4,0.2)$. The rescaled curves, obtained from the dashed curves with the given $M_0$, match the dotted curves with $M_0^{'}=U^{\rm MF}$ after the transients. Curves with $M_0^{'}=U^{\rm MF}$ are compared with those of $M_0^{'}=b^{z^{\rm MF}/2}M_0$ for (c) $M_0=0.5$ ($b=2$) and (d) $M_0=0.2$ ($b=4$). Each rescaled curve matches the dotted curve of $M_0^{'}=U^{\rm MF}$, rather than the chain curve of $M_0^{'}=b^{z^{\rm MF}/2}M_0$ after a microscopic time-scale. Double-logarithmic scales are used.}
\end{figure}

We now verify the scaling properties described by Eq.~(\ref{SCYYf1}). To this end, we solve numerically Eq.~(\ref{dmfhamil}) with an arbitrary $M_0$ satisfying $M_0<b^{-z^{\rm MF}}\sqrt{2b^{z^{\rm MF}}-1}$. Then the evolution curve of $M$ is rescaled according to Eq.~(\ref{gscf}). The rescaled curve is compared with the one beginning with $U^{\rm MF}(b,M_0)$. From Fig.~\ref{MFNR} (a) and (b), one can find that these two curves collapse onto each other in both short-time and long-time stage after a microscopic time scale. In addition, Fig.~\ref{MFNR} (c) and (d) show the necessity of the universal characteristic function. These results confirm the scaling theory characterized by the universal characteristic function, $U^{\rm MF}$, in the mean-field theory.

\begin{figure}
  \centerline{\epsfig{file=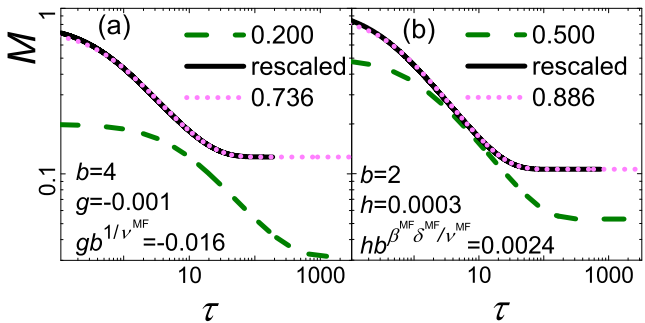,width=1.0\columnwidth}}
  \caption{\label{MFHG} (Color online) Independence of $U^{\rm MF}$ on (a) $g$ and (b) $h$. The rescaled curve match the curve beginning with $U^{\rm MF}$, which is obtained at $g=0$ and $h=0$ according to Eq.~(\ref{SCYYf1}).}
\end{figure}

Then we confirm that the universal characteristic function, $U^{\rm MF}$, is independent of $g$ and $h$. For this purpose, we numerically solve Eq.~(\ref{dmfhamil}) in presence of finite $g$ and $h$. Figure~\ref{MFHG} shows that the rescaled curve and the curve from $U^{\rm MF}(b,M_0)$, determined at $g=0$ and $h=0$, collapse onto each other. Therefore, $U^{\rm MF}(b,M_0)$ depends only on $M_0$ and $b$, confirming the discussions in Sec.~\ref{scalingt}.

\subsubsection{Universality of $U^{\rm MF}$}
\begin{figure}
  \centerline{\epsfig{file=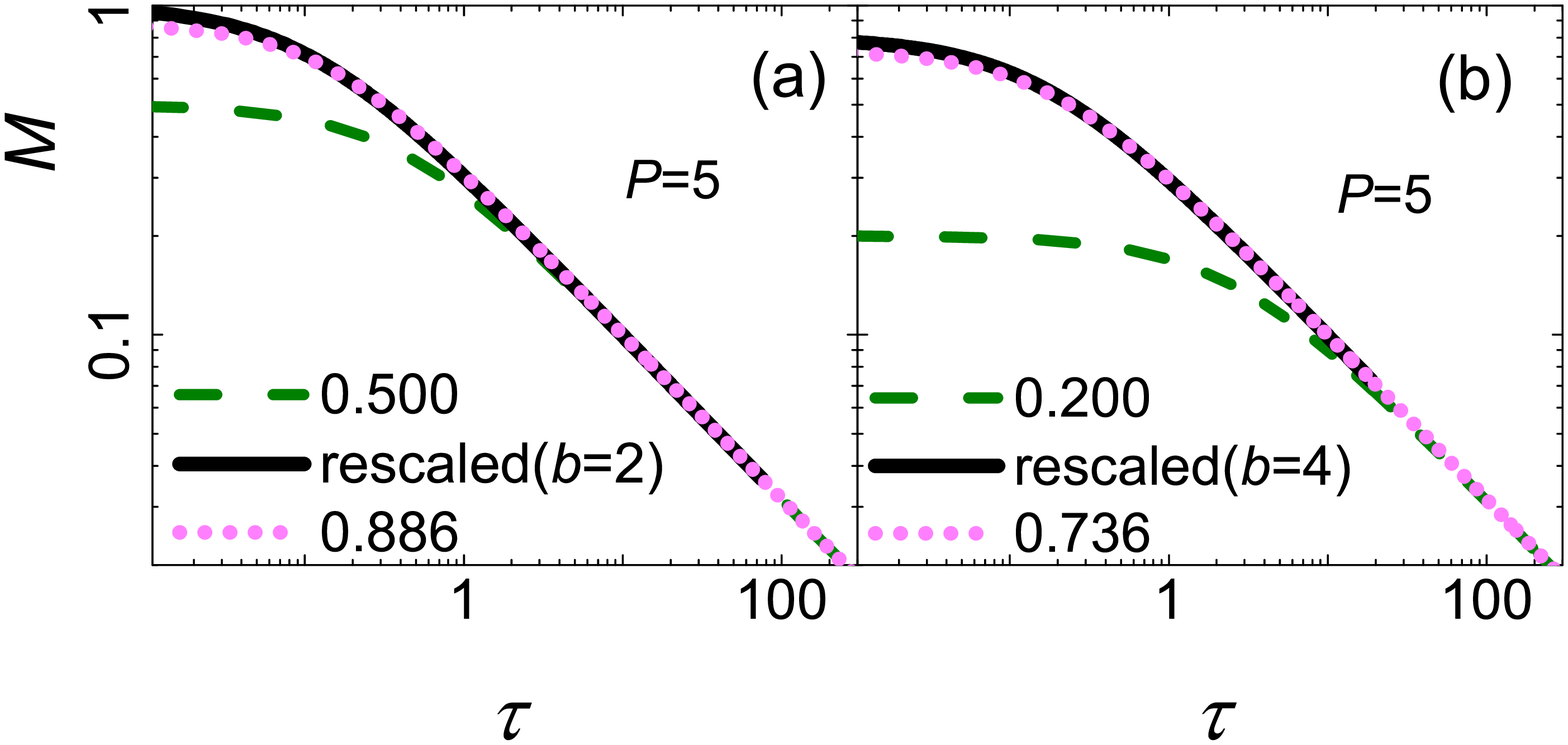,width=1.0\columnwidth}}
  \caption{\label{UMFA} (Color online) Verification of the universality of the universal characteristic function for (a) $U^{\rm MF}(2,0.5)$ and (b) $U^{\rm MF}(4,0.2)$ with $P=5$ and $h_x=h_{xc}^{\rm MF}=5$. The rescaled curves, obtained from the dashed curves with the given $M_0$, match the dotted curves with $M_0^{'}=U^{\rm MF}$ after the transients.}
\end{figure}

To investigate the universality of $U^{\rm MF}$, we study the mean-field theory of the multi-leg quantum Ising ladders. The mean-field Hamiltonian of the quantum Ising ladder model (\ref{HIsingl}) is
\begin{equation}
\tilde{H}_{\rm MF}=-3M\sigma_z-h_x\sigma_x. \label{mfhamill}
\end{equation}
Compared to Eq.~(\ref{mfhamil}), we see that the figure of the first term is just the number of bonds connecting to each spin. We can thus generalize this ladder model to a series of multi-leg ladder models, each of which consists of one Ising chain coupled with $P-2$ mutually independent other chains. The dynamical equation describing the evolution of the magnetization is then
\begin{equation}
\frac{dM}{d\tau}=2PM-2PM^3-2h_x M\sqrt{1-M^2}. \label{dmfhamil1}
\end{equation}
It can be readily checked that the dynamics of all the multi-leg ladders described by Eq.~(\ref{dmfhamil1}) belong to the same universality class as that by Eq.~(\ref{dmfhamil}). Moreover, the condition, $dM/d\tau=0$ when $M=1$, makes the order parameter bounded and the saturated magnetization is also $M=1$. At the critical point $h^{\rm MF}_{xc}=P$, the universal solution then approximately satisfies
\begin{equation}
Y(M)=4P\tau+Y(M_0), \label{YPfun}
\end{equation}
Comparing Eq.~(\ref{YPfun}) with Eq.~(\ref{YYfun}), one can readily find that only the coefficient of $\tau$ is different. However, according to Eq.~(\ref{SCYYf}), this difference does not change the form of $U^{\rm MF}$. Therefore, the characteristic function of Eq.~(\ref{dmfhamil1}) is also Eq.~(\ref{SCYYf1}). This is verified by solving Eq.~(\ref{dmfhamil1}) numerically with various initial $M_0$ for an arbitrary $P$ in Fig.~\ref{UMFA}, in which the curve, beginning with $M_0^{'}=U^{\rm MF}(b,M_0)$, matches the rescaled curve from $M_0$ perfectly. This confirms the universality of the characteristic function in the mean-field theory.

\subsection{\label{scaling}Numerical results via ITEBD}
In Sec.~\ref{MF}, universal scaling behavior has been confirmed in the mean-field theory through the approximate analytic solution. When quantum fluctuations are taken into account, however, we fail to obtain such an analytic form, as it is difficult to expand the initial state in the eigenstates of the Hamiltonian analytically. So, the dynamic scaling properties, described by the characteristic function, are needed to be examined numerically. To further study the universal short-imaginary-time dynamics, in this section, we simulate the imaginary-time evolution using the ITEBD algorithm.

\begin{figure}
  \centerline{\epsfig{file=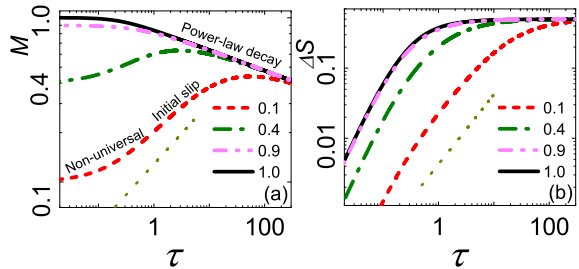,width=1.0\columnwidth}}
  \caption{\label{farbm} (Color online) The imaginary-time evolutions of (a) $M$ and (b) $\Delta S$ with various $M_0$ indicated at $g=0$ and $h=0$. The thin dotted lines in (a) and (b) are the lines of $M\sim M_0\tau^\theta$ and $\Delta S\sim M_0^2\tau^{2x_0/z}$, respectively, for vanishing small $M_0$. Double-logarithmic scales are used in both figures.}
\end{figure}

Figure \ref{farbm} (a) shows the imaginary-time evolution of the order parameter $M$ with different initial magnetization $M_0$ for the quantum Ising model (\ref{HIsing}). The initial-slip stage, in which $M$ increases, shrinks as $M_0$ increases. Then, $M$ enters the power-law decay stage, in which $M\sim \tau^{-\beta/\nu z}$, independent of the value of $M_0$. These features are consistent with those with very small $M_0$. However, in the initial slip stage, the shapes of the curves of $M$ versus $\tau$ change with $M_0$. This is different from the situation for very small $M_0$, for which the curves in the initial-slip stage are straight lines with identical slopes on a double-logarithmic scale \cite{yinprb}. This indicates the necessity of the presence of the universal characteristic function $U$. The evolution of entanglement entropy-difference $\Delta S$ is shown in Fig. \ref{farbm} (b). $\Delta S$ saturates at a constant independent of $M_0$ for long times. In the early stage, the slopes for larger $M_0$ are different from those for small $M_0$. This also indicates the necessity of $U$.

\subsubsection{Determination of $U(b,M_0)$}
To determine numerically the form of $U$, according to Eq. (\ref{gsc}), we first let the system evolve at a certain $g$ and $h$ starting from an initial magnetization $M_0$, and rescale $M$ and $\tau$ by $M\rightarrow b^{\beta/\nu}M$ and $\tau\rightarrow b^{-z}\tau$, respectively. Then we run the system at $b^{1/\nu}g$ and $b^{\beta\delta/\nu}h$ starting from a series of $M'_0$, and compare each of these curves with the rescaled curve to find the one that most fit it. The corresponding initial magnetization thus satisfies $M'_0=U(b,M_0)$.

\begin{figure}
  \centerline{\epsfig{file=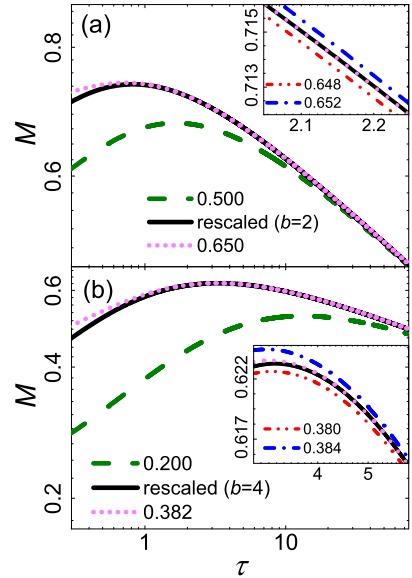,width=0.75\columnwidth}}
  \caption{\label{fmcp} (Color online) Determinations of (a) $U(2,0.5)$ and (b) $U(4,0.2)$. The rescaled curves, which are obtained from the dashed curves with the given $M_0$, match the dotted curves with $M_0^{'}=U$ after the transients. Both insets show two other evolution curves with $M_0$ indicated near $M_0^{'}=U$. Double-logarithmic scales are used.}
\end{figure}

The generalized scale transformation (\ref{gsc}) show that $U$ is independent of the choice of $g$ and $h$. So, we work at the critical point, $g=0$ and $h=0$, for simplicity. Figure~\ref{fmcp} shows the determination of $U$ for two sets of values. It is clear that after the non-universal microscopic time, the curves starting with $U(b,M_0)$ fit well with the rescaled curves, in both the short- and long-time stages. In contrast, if the initial magnetization is not $U(b,M_0)$, the curves of the subsequent evolution deviate from the rescaled curves as can be seen in the insets.

\begin{figure}
  \centerline{\epsfig{file=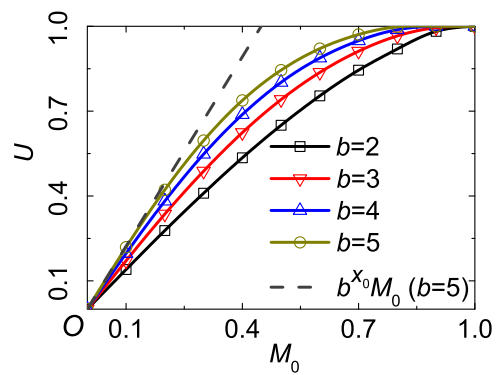,width=0.75\columnwidth}}
  \caption{\label{fchi} (Color online) The measured characteristic function for several $b$. The curves are spline interpolation between data denoted by symbols. The straight dashed line represents the power law $b^{x_0}M_0$ for $b=5$ and $x_0=0.498$ (valid when $M_0$ is small) \cite{yinprb}.}
\end{figure}

$U(b,M_0)$ measured for model (\ref{HIsing}) is shown in Fig.~\ref{fchi}. Two fixed points $U(b,0)=0$ and $U(b,1)=1$ are manifest for different rescaling factors $b$. Between the two fixed points, $U$ increases as $M_0$ increases. Comparing the line of $b^{x_0}M_0$ for $b=5$ with $U(5,M_0)$, one finds that they fit well when $M_0$ is small and deviates from each other when $M_0$ grows larger. $U$ also increases as $b$ increases. For larger $b$, the curve deviates from the power-law form for smaller $M_0$, and tends to saturate at smaller $M_0$.

\begin{figure}
  \centerline{\epsfig{file=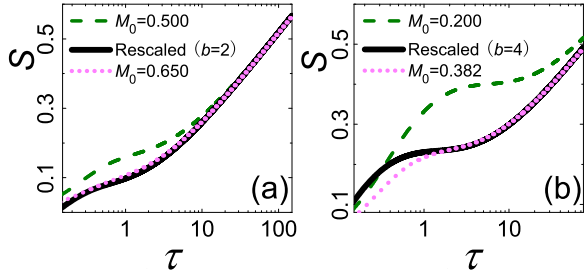,width=1.0\columnwidth}}
  \caption{\label{fscp} (Color online) (a) $U(2,0.5)=0.650$ and (b) $U(4,0.2)=0.382$ checked with $S$ at the critical point.
  Semi-logarithmic scales are used.}
\end{figure}

We can apply the scale transformation of the entropy $S$, Eq.~(\ref{eesc}), to determine $U$ as well. However, we apply this to check the foregoing determined $U$. According to Eq. (\ref{eesc}), the scale-transformed curve $S(b^z\tau,g,h,M_0)-(c/6)\textrm{log}b$ which starts with $M_0$ should coincide with the curve of $S(\tau,b^{1/\nu}g,b^{\beta\delta/\nu}h,U(b,M_0))$. From Fig.~\ref{fscp}, it can be seen that, after a microscopic time scale, the curve from $M_0^{'}=U(b,M_0)$, determined in Fig.~\ref{fmcp}, and the corresponding rescaled curve collapse onto each other. As the entanglement entropy $S$ is closely related with the correlation length $\xi$, which is responsible to the universal behavior in the critical region, we expect that the universal characteristic function $U$ must be considered for all the macroscopic quantities.


\subsubsection{Off-critical-point situation}
When the system deviates a little from the critical point, i.e., $g\neq 0$ or $h\neq 0$ (or both), the generalized scale transformation (\ref{gsc}) still holds, and the characteristic function is not dependent on the value of $g$ or $h$ as long as the system is close to the critical point. We have demonstrate this independence in the mean-field approximation in Sec.~\ref{MF}. In this section, we show that this scaling property still holds in presence of the quantum fluctuations.

\begin{figure}
  \centerline{\epsfig{file=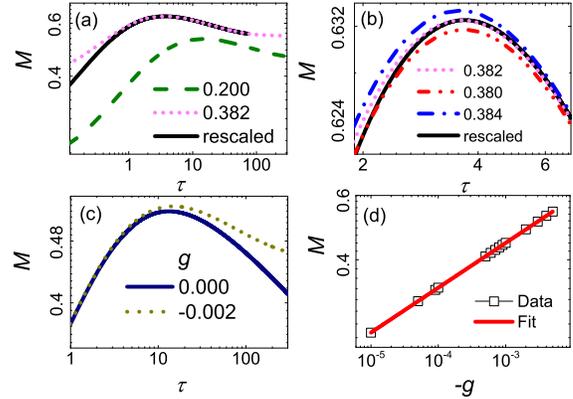,width=1.0\columnwidth}}
  \caption{\label{fmg} (Color online) $U(4,0.2)=0.382$ determined from $M$ when $g=-0.002$ in (a) and (b). The curve with $M_0=0.2$ for $g=-0.002$ is compared with that for $g=0$ in (c). At long times, the magnetization for $g=-0.002$ tends to its equilibrium value, which satisfies $M\sim (-g)^{\beta}$. Fitting of the equilibrium $M$ versus various $g$ in (d) gives $\beta=0.123$, which is close to its exact value $1/8$.  Double-logarithmic scales are used.}
\end{figure}

To this end, first we study a system slightly deviated from the critical point with $g\equiv h_x-h_{x\textrm{c}}\neq0$. When $g>0$, the system is in the disordered phase and $M$ tends to $0$ as the evolution continues; when $g<0$, the system is in the ordered phase and $M$ tends to $M\propto (-g)^\beta$ in equilibrium as shown in Fig.~\ref{fmg} (c) and (d). From Fig.~\ref{fmg} (a) and (b), one can find that the value $M_0^{'}$, from which the evolution curve and the rescaled curve from $M_0$ collapse, is almost the same as $U(b,M_0)$ determined at the critical point, although the shape of the curve for finite $g$ is significantly different from that for $g=0$, as displayed in Fig.~\ref{fmg} (c).

\begin{figure}
  \centerline{\epsfig{file=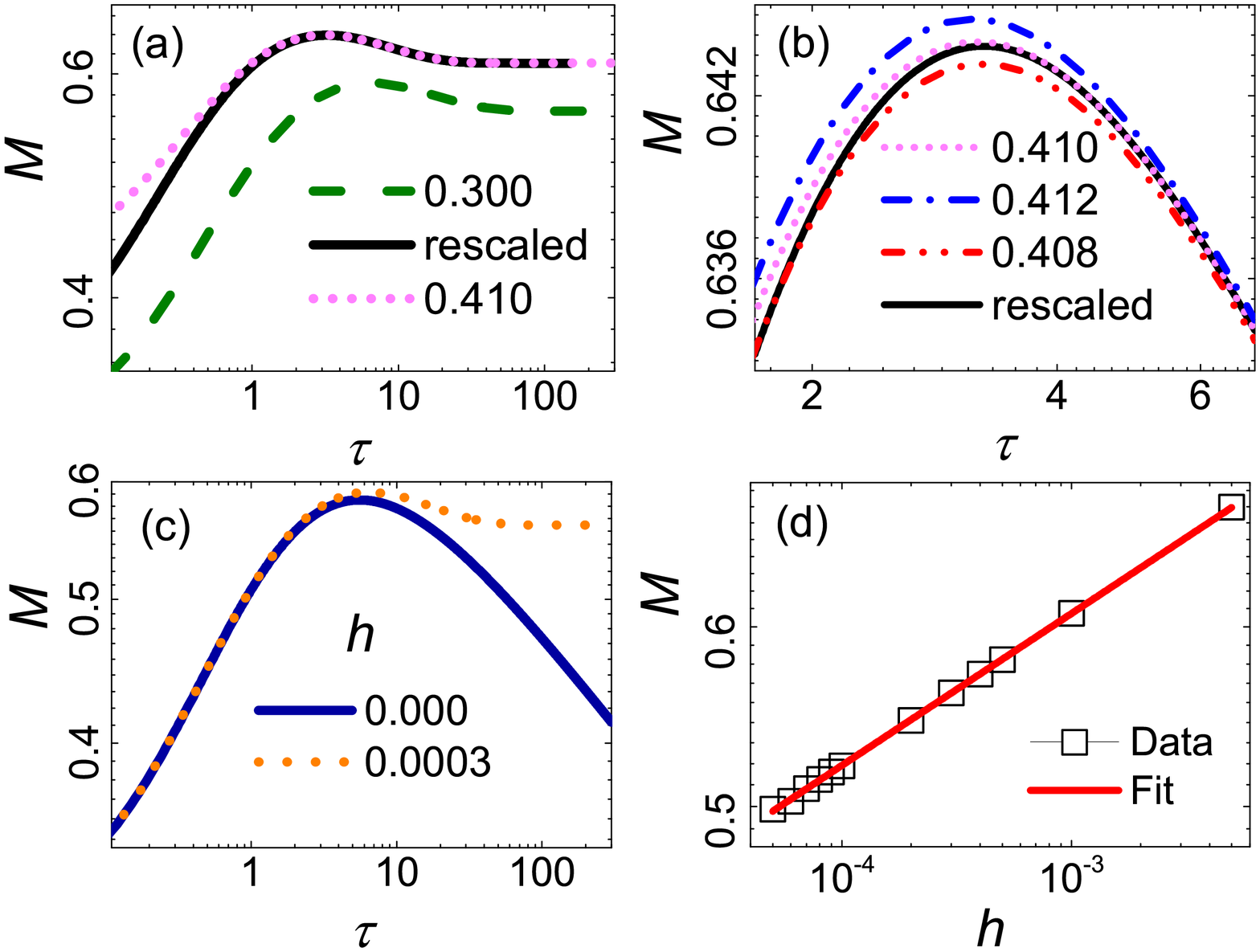,width=1.0\columnwidth}}
  \caption{\label{fmh} (Color online) $U(2,0.3)=0.410$ determined from $M$ when $h=0.0003$ in (a) and (b). The curve with $M_0=0.3$ for $h=0.0003$ is compared with that for $h=0$ in (c). At long times, the magnetization for $h=0.0003$ tends to its equilibrium value, which satisfies $M\sim h^{1/\delta}$. Fitting of the equilibrium $M$ versus various $h$ in (d) gives $1/\delta=0.0668$, which is close to its exact value $1/15$.  Double-logarithmic scales are used.}
\end{figure}

Then we examine the situation with a small symmetry-breaking field $h$. We will restrict it to be in the same direction of the initial order parameter in order to prevent $M$ from dropping below zero in the evolution. In equilibrium, $M$ should tend to $M\propto h^{1/\delta}$ as shown in Fig.~\ref{fmh} (c) and (d). In spite of the fact that the curve of $h\neq 0$ is different from that of $h=0$, as displayed in Fig.~\ref{fmh} (c),  Fig.~\ref{fmh} (a) and (b) show that $U(b,M_0)$ is identical with the corresponding one in the absence of $h$.

We have measured the dependence of $U$ on $b$ and $M_0$ in presence of $g$ and $h$. $b$ ranges from $2$ to $5$ with an increment $1$ and $M_0$ ranges from $0.1$ to $0.9$ with an increment $0.1$. Part of the results is listed in Table \ref{tab1}. From this table, we find that the characteristic function $U(b,M_0)$ is really independent of $g$ and $h$. Moreover, we have also measured the deviation of the measured $M_0^{'}$ from $U(b,M_0)$, which is determined at $g=0$ and $h=0$, for various $g$ and $h$ in different magnitudes. The results, listed in Table \ref{tab2}, show that the deviations are almost all zero in the present degree of precision and demonstrate that the universal characteristic function is independent of $g$ and $h$.
\begin{table}[htbp]
  \centering
  \caption{$U(b,M_0)$ at and off the critical point}
    \begin{tabular}{c|l| c c c c c}
    \hline
    \hline
    \multicolumn{2}{c|}{$M_0$} & 0.1   & 0.3   & 0.5   & 0.7   & 0.9  \\
    \hline
    $U(2,M_0)$ & $g=h=0$ & 0.140  & 0.410  & 0.650  & 0.845  & 0.980  \\
          & $h=0.0001$ & 0.140  & 0.410  & 0.650  & 0.845  & 0.979  \\
          & $g=0.001$   & 0.141  & 0.410  & 0.650  & 0.845  & 0.980  \\
          & $g=-0.001$   & 0.141  & 0.411  & 0.651  & 0.845  & 0.980  \\
    \hline
    $U(4,M_0)$ & $g=h=0$ & 0.197  & 0.548  & 0.802  & 0.949  & 1.000  \\
          & $h=0.0001$ & 0.196  & 0.547  & 0.802  & 0.949  & 0.999  \\
          & $g=0.001$   & 0.197  & 0.547  & 0.800  & 0.948  & 0.999  \\
          & $g=-0.001$   & 0.197  & 0.548  & 0.802  & 0.950  & 1.000  \\
    \hline
    \end{tabular}%
  \label{tab1}%
\end{table}%

\begin{table}[htbp]
  \centering
  \caption{Deviation of $M_0^{'}$ from $U$ off the critical point}
    \begin{tabular}{c| c c c c c}
    \hline
    \hline
    \multicolumn{1}{c|}{$h$} &0.00001 & 0.00005   & 0.0001   & 0.0005   & 0.001  \\
    \hline
      $M_0^{'}-U(2,0.3)$   &0.000 & 0.000  & 0.000  & 0.000  & 0.001  \\
    \hline
    \multicolumn{1}{c|}{$-g$} &0.00001 & 0.00005   & 0.0001   & 0.0005   & 0.001  \\
    \hline
      $M_0^{'}-U(2,0.3)$   &0.000 & 0.000 & -0.001  & 0.001  & 0.000 \\
    \hline
    \multicolumn{1}{c|}{$g$} &0.00001 & 0.00005   & 0.0001   & 0.0005   & 0.001  \\
    \hline
      $M_0^{'}-U(2,0.3)$   &0.000 & 0.000 & 0.000  & 0.000  & 0.000 \\
    \hline
    \end{tabular}%
  \label{tab2}%
\end{table}%


\subsubsection{Universality of $U(b,M_0)$}
In this section, we confirm the universality of the characteristic function. Firstly, we show that the universal characteristic function only depends on the value of the initial magnetization, rather than its detailed realization. Secondly, the universality of the characteristic function is confirmed for other models in the same universality class.

\begin{figure}
  \centerline{\epsfig{file=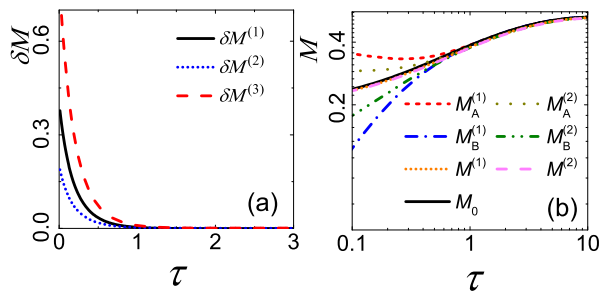,width=0.85\columnwidth}}
  \caption{\label{initial} (Color online) (a) Evolution of the difference of the magnetization between sublattice $A$ and $B$ with $M_0=0.2$. For different realizations of the initial magnetization, $\delta M$ becomes vanishing after a microscopic time-scale. (b) Imaginary-time evolution of the magnetizations in sublattice $A$ and $B$ and their averages, $M^{(1)}$ and $M^{(2)}$ for different realizations of the initial magnetization. All the curves converge to the solid curve of the homogeneous initial condition after the transients. The different initial realizations in (a) and (b) are defined as $M^{(1)}_{0A}=0.4$, $M^{(1)}_{0B}=0.0$, $M^{(2)}_{0A}=0.3$, $M^{(1)}_{0B}=0.1$, $M^{(3)}_{0A}=0.5$ and $M^{(3)}_{0B}=-0.1$. As $M^{(3)}_{0B}$ is negative, the last one is absent in (b), where double-logarithmic scales are used.}
\end{figure}

To explore the effects induced by the initial state, we use the staggered initial wave function (\ref{istate}) and define the magnetization-difference of different sublattices, $\delta M$, as
\begin{equation}
\delta M\equiv |M_A-M_B|, \label{dmamb}
\end{equation}
where $M_A$ and $M_B$ are the magnetization for sublattice $A$ and $B$, respectively. Figure \ref{initial} (a) shows that $\delta M$ approaches zero after the transients, irrespective of the realizations of the initial magnetization. This indicates the high-momentum modes decay in a microscopic time-scale. After that, the long-wavelength modes control the universal behavior, described by the universal characteristic function. These long-wavelength modes only depend on the averaged initial magnetization $M_0$. These remarks are further illustrated in Fig.~\ref{initial} (b), which shows that the local order parameters of different sublattices and their averages tend to a unified curve, which is identical to the one starting with the homogeneous realization of $M_0$. These results are consistent with those in the previous study \cite{yinprb}, demonstrating that the universal critical initial-slip behavior is independent of the detailed realization of the initial magnetization.

\begin{figure}
  \centerline{\epsfig{file=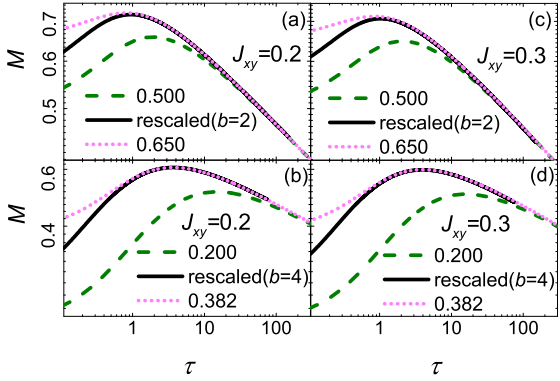,width=1.0\columnwidth}}
  \caption{\label{xxzf} (Color online) $U(2,0.5)=0.650$ and $U(4,0.2)=0.382$ checked in the quantum XXZ model for $J_{xy}=0.2$ and $J_{xy}=0.3$. Double-logarithmic scales are used.}
\end{figure}

\begin{figure}
  \centerline{\epsfig{file=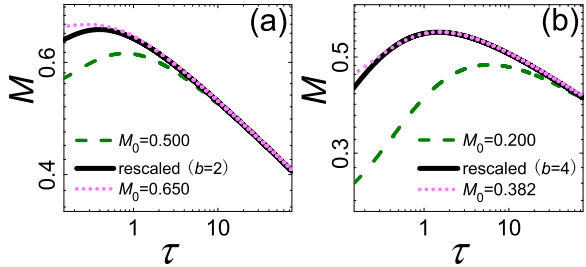,width=1.0\columnwidth}}
  \caption{\label{fladm} (Color online) (a) $U(2,0.5)=0.650$ and (b) $U(4,0.2)=0.382$ checked in the quantum Ising-ladder at its critical point.
   Double-logarithmic scales are used.}
\end{figure}

Figure \ref{xxzf} shows the imaginary-time evolution for the quantum XXZ model~(\ref{HXXZ}) for $J_{xy}=0.2$ and $0.3$ at their respective critical points. From $M_0^{'}=U(b,M_0)$, where $U(b,M_0)$ is chosen to be identical with the value of model (\ref{HIsing}), the curve and the corresponding rescaled curve from $M_0$ collapse onto each other after the early microscopic time stage. Although the shape of the evolution curve in this model is slightly different from that in the $1$D Ising chain, the values of $U$ are identical for both models, thus confirming the the universality of the characteristic function. Similarly, Fig.~\ref{fladm} confirms the universality of $U(b,M_0)$ in the quantum Ising-ladder model~(\ref{HIsingl}). These results demonstrate that the universal characteristic function depends only on the universality class.

\section{\label{summary}Summary}
We have studied in this paper the relaxation quantum critical dynamics in imaginary time with an arbitrary initial order parameter. We have shown that a universal characteristic function $U$ must be introduced to describe the universal properties in both short and long times for an arbitrary initial magnetization $M_0$ similar to the classical case. This characteristic function contains the rescaling factor $b$ and $M_0$ as its arguments. It is identical for the models belonging to one universality class. According to the scale transformation including $U$, the form of this function has been determined for the $1$D transverse-field Ising model in a mean-field theory and with numerical results. The value of the characteristic function is shown to be independent of the detailed realization of the initial magnetization. The universality of $U$ has also been confirmed in the transverse-field quantum XXZ model and the quantum Ising-ladder model. Although we have only studied one universality class, the theory should be applicable to models in other universality classes as well.

\section*{Acknowledgements}
We wish to thank Zhibing Li for his discussions. This project was supported by NNSFC (10625420).

\end{document}